\documentclass{elsart3-1}

\usepackage{amssymb,epsfig}

\usepackage[english,francais]{babel}

\newtheorem{e-proposition}[theorem]{Proposition}

\newtheorem{e-definition}[theorem]{Definition\rm}

\renewcommand{\theequation}{\arabic{equation}}
\newcommand{\vct}[1]{{\mbox {\boldmath $#1$}}}

\def\build#1_#2^#3{\mathrel{\mathop{\kern 0pt#1}\limits_{#2}^{#3}}}

\usepackage{epsfig,amsmath,amssymb}

\def\og{\leavevmode\raise.3ex\hbox{$\scriptscriptstyle\langle\!\langle$~}}
\def\fg{\leavevmode\raise.3ex\hbox{~$\!\scriptscriptstyle\,\rangle\!\rangle$}}

\begin{document}

\begin{frontmatter}


\selectlanguage{english}
\title{Intermittency and universality in a Lagrangian model of velocity gradients in three-dimensional turbulence}


\selectlanguage{english}
\author[authorlabel1]{Laurent Chevillard},
\ead{chevillard@jhu.edu}
\author[authorlabel1]{Charles Meneveau}
\ead{meneveau@jhu.edu}

\address[authorlabel1]{Department of Mechanical Engineering, the Johns
Hopkins University, 3400 N. Charles Street, Baltimore, MD, USA.}


\medskip
\begin{center}
{\small Received *****; accepted after revision +++++}
\end{center}

\begin{abstract}
The universality of intermittency in hydrodynamic turbulence is
considered based on a recent model for the velocity gradient
tensor evolution. Three possible versions of the model are
investigated differing in the assumed correlation time-scale and
forcing strength. Numerical tests show that the same (universal)
anomalous relative scaling exponents are obtained for the three
model variants. It is also found that transverse velocity
gradients are more intermittent than longitudinal ones, whereas
dissipation and enstrophy scale with the same exponents. The
results are consistent with the universality of intermittency and
relative scaling exponents, and suggest that these are dictated by
the self-stretching terms that are the same in each variant of the
model.

\vskip 0.5\baselineskip

\selectlanguage{francais} \noindent{\bf R\'esum\'e} \vskip
0.5\baselineskip \noindent {\bf Intermittence et universalit\'e
d'un mod\`ele lagrangien des gradients de vitesse en turbulence
3D.} Le caract\`ere universel du ph\'enom\`ene d'intermittence en
turbulence est \'etudi\'e \`a partir d'un mod\`ele r\'ecent
r\'egissant l'\'evolution du tenseur des gradients de vitesse.
Trois versions possibles du mod\`ele, pour lesquelles les
hypoth\`eses retenues pour le temps de corr\'elation et
l'amplitude du for\c{c}age sont diff\'erentes, sont analys\'ees.
Une int\'egration num\'erique des \'equations montre que les
exposants anormaux des moments relatifs sont les m\^emes pour les
trois variantes du mod\`ele. Il est de plus montr\'e que les
gradients transversaux de vitesse sont plus intermittents que les
longitudinaux alors que la dissipation et l'enstrophie se
comportent comme des lois de puissance de m\^eme exposant. Ces
r\'esultats sont coh\'erents avec l'universalit\'e des exposants
relatifs et sugg\`erent l'importance du terme d'auto-\'etirement,
qui est identique dans les trois variantes du mod\`ele. {\it Pour
citer cet article~: L. Chevillard, C. Meneveau, C. R. Mecanique
\textbf{335} (2007).}

 \vskip
0.5\baselineskip \noindent{\small{\it Mots-cl\'es~:} Turbulence;
Intermittence~; G\'eom\'etrie}}
\end{abstract}
\end{frontmatter}

\selectlanguage{francais}

\selectlanguage{english}
\section{Introduction}
\label{}




Progress in understanding the small-scale structure of
three-dimensional turbulent flow requires the study of the
velocity gradient tensor $A_{ij} = \partial u_i/\partial x_j$,
where \textbf{u} denotes the velocity vector. In incompressible
flow, $\textbf{A}$ is trace-free, i.e. $A_{ii}=0$. The dynamical
evolution  of $\textbf{A}$ is obtained by taking the gradient of
the Navier-Stokes equation:
\begin{equation}\label{eq:NS}
\frac{dA_{ij}}{dt} = -A_{ik}A_{kj}-\frac{\partial^2p}{\partial
x_i\partial x_j}+\nu\frac{\partial^2A_{ij}}{\partial x_k\partial
x_k}\mbox{ ,}
\end{equation}
where $d/dt$ stands for the Lagrangian material derivative, $p$ is
the pressure divided by the density of the fluid and $\nu$ is the
kinematic viscosity.  Neglecting viscous effects and the
anisotropic part of the pressure Hessian entering in Eq.
(\ref{eq:NS}) leads to a closed formulation of the dynamics of the
velocity gradient tensor known as the Restricted-Euler (RE)
equations \cite{Vieille2,Cant1}. RE equations predict several
phenomena observed in various experimental \cite{BosTao02,Zef03}
and numerical \cite{Cant2} studies of turbulence, such as
preferential alignments of vorticity and preferential axisymmetric
extension. Recently, a system of differential equations describing
longitudinal and transverse velocity increments has been derived
from this approximation and predicts non-Gaussian statistics (and
in particular skewness) of the components of \textbf{A}
\cite{LiMen06}. However, in this system as well as in the RE
equations, the neglect of pressure Hessian and viscous effects
leads to singularities and precludes the establishment  of
stationary statistics due to undamped  effects of the
self-streching term. To adress this deficiency of RE dynamics, and
based on prior works \cite{GirPop90,ChePum99,JeoGir03}, a new
model has been proposed \cite{CheMen06}: the Recent Fluid
Deformation (RFD) closure. It models both pressure Hessian and
viscous term entering in Eq. (\ref{eq:NS}). The RFD closure is
based on the dynamics and the geometry of the deformation
experienced by the fluid during its most recent evolution. A
de-correlation time scale $\tau$ entering the various closure
terms has to be specified, as well as a Gaussian forcing term. It
was shown \cite{CheMen06} that the system with $\tau$ chosen equal
to the Kolmogorov scale and a fixed Gaussian forcing amplitude,
reproduces stationary statistics with a number of geometric
features of the velocity gradient, as well as relative scaling
exponents of high-order moments,  in strikingly close agreement to
experimental and numerical measurements for real Navier-Stokes
turbulence.

In order to explore the possible universality properties of the
model, in this Letter we study two different choices for the
time-scale and also explore the consequences of varying  the
amplitude of the forcing term. We pose the question of whether the
anomalous scaling exponents vary from case to case, and also
extend the analysis to include scaling properties of the
dissipation.

The model proposed by Ref. \cite{CheMen06} begins with a change of
variables, expressing the pressure in terms of the Lagrangian
coordinates $\textbf{X}$. One may define a mapping
$\mathcal{M}_{t_0,t}$ between Eulerian and Lagrangian coordinates:
$\mathcal{M}_{t_0,t}:\textbf{X}\in \mathbb R^3\mapsto
\textbf{x}\in \mathbb R^3$, where $\textbf{x}(\textbf{X},t)$
denotes the position at a time $t$ of a fluid particle which was
at the position $\textbf{x}(\textbf{X},t_0)=\textbf{X}$ at the
initial time $t_0$. The Jacobian matrix of the inverse mapping
obeys $d/dt(\partial X_p/\partial x_i)=-A_{ki}(\partial
X_p/\partial x_k)$. As argued before \cite{CheMen06}, for a
relative short period of time (typically when $t-t_0=\tau$, where
$\tau$ is a characteristic Lagrangian decorrelation time scale for
the velocity gradient tensor), the solution can be approximated as
the matrix exponential of the velocity gradient itself, namely
$(\partial X_p/\partial x_i) = \left(\exp[-\tau {\bf
A}]\right)_{pi}$.

With an Eulerian-Lagrangian change of variables,  the pressure
Hessian can be written in the following way:
\begin{equation}\label{eq:ChangVar}
\frac{\partial ^2p(t)}{\partial x_i\partial x_j} \approx
\frac{\partial X_p}{\partial x_i}\frac{\partial X_q}{\partial
x_j}\frac{\partial ^2p(t)}{\partial X_p\partial X_q}\approx -
\frac{\mbox{Tr}(\textbf{A}^2)}{\mbox{Tr}
(\textbf{C}_\tau^{-1})}\left(\textbf{C}_\tau^{-1}\right)_{ij}
\mbox{ .}
\end{equation}
In the first approximation, spatial gradients of $(\partial
X_p/\partial x_i)$ have been neglected \cite{CheMen06}. In the
second approximation, the short-time solution for $(\partial
X_p/\partial x_i)$ mentioned before, the isotropy assumption for
the Lagrangian pressure Hessian ($\partial^2 p/\partial
X_p\partial X_q \sim \delta_{pq}$), and the trace-free condition
of $A_{ij}$ have been used. Moreover, $\textbf{C}_\tau$ is the
short-time Cauchy-Green tensor \cite{CheMen06}: $\textbf{C}_\tau =
e^{\tau \textbf{A}}e^{\tau \textbf{A}^T}$. This model (Eq.
(\ref{eq:ChangVar})) can be viewed as a local version of the
``tetrad model'' \cite{ChePum99}. The time-scale $\tau$ is a model
parameter and can be viewed as a characteristic time scale of the
small dissipative scales of turbulence. Recent material
deformation history can also be used to model other Hessian
tensors entering in the Navier-Stokes equation (Eq. (\ref{eq:NS}))
such as the viscous part. The resulting Hessian of $\textbf{A}$ is
modeled as a friction term and the characteristic time scale
entering in the description is given by the (Reynolds number
independent) integral time scale $T$ \cite{CheMen06}:
\begin{equation}\label{eq:ModViscA}
\nu \nabla^2 \textbf{A}  \approx - \frac{1}{T}
\frac{\mbox{Tr}(\textbf{C}_\tau^{-1})}{3}\textbf{A}\mbox{ ,}
\end{equation}
which is a stationary version of the model of Jeong and Girimaji
\cite{JeoGir03}. Finally, combining Eqs. (\ref{eq:ChangVar}) and
(\ref{eq:ModViscA}) into Eq. (\ref{eq:NS}), one obtains a model
for the dynamic evolution of $\textbf{A}$ along a Lagrangian
trajectory \cite{CheMen06},
\begin{equation}\label{eq:Determourmodel}
d\textbf{A} =  \left(-\textbf{A}^2+
\frac{\mbox{Tr}(\textbf{A}^2)}{\mbox{Tr}(\textbf{C}_{\tau}^{-1})}
\textbf{C}_{\tau}^{-1}
-\frac{1}{T}\frac{\mbox{Tr}(\textbf{C}_{\tau}^{-1})}{3} \textbf{A}\right)dt
+d\textbf{W}\mbox{ .}
\end{equation}
A stochastic forcing term $d\textbf{W}$ has been added to model
the combined action of large-scale forcing and neighboring eddies.
The time evolution of $\textbf{A}$ (Eq. (\ref{eq:Determourmodel}))
is thus given by eight independent coupled ordinary (or stochastic
depending on the forcing $d\textbf{W}$) differential equations.

In order to understand the roles played by the pressure Hessian
(Eq. (\ref{eq:ChangVar})) and the viscous term (Eq.
(\ref{eq:ModViscA}))  in Eq. (\ref{eq:Determourmodel}), some
analysis can be carried out. Doing so for arbitrary initial
conditions $\textbf{A}(t_0)$ is difficult analytically because of
the high dimension of the phase space. However, following Ref.
\cite{ChePum99} one may consider the  decaying case (with
$d\textbf{W}=0$) along a particular direction corresponding to
strain with two equal positive, and one negative, eigenvalues.
Along this direction on  the  'Vieillefosse tail', the tensor
$\textbf{A}$ and the evolution of the relevant eigenvalue
$\lambda(t)$ from the model (Eq. (\ref{eq:Determourmodel})) are
given by
\begin{equation}\label{eq:InitVieille}
\textbf{A} = \begin{pmatrix}
  \lambda(t) & 0 & 0 \\
  0 & \lambda(t) & 0 \\
  0 & 0 & -2\lambda(t)
\end{pmatrix}\build{\Rightarrow}_{}^{d\textbf{W}\equiv 0}\frac{d\lambda}{dt} = \frac{4e^{-2\tau\lambda}
-e^{4\tau\lambda}}{2e^{-2\tau\lambda}+e^{4\tau\lambda}}\lambda^2 -
\frac{2e^{-2\tau\lambda}+e^{4\tau\lambda}}{3T}\lambda \mbox{ .}
\end{equation}
The solution of the ODE in Eq. (\ref{eq:InitVieille}) is such that
$\lambda(t)$ retains the same sign as $\lambda(0)$. Let us also
recall that in the RE system, i.e. $d\textbf{A}/dt =
-\textbf{A}^2+\textbf{I}\mbox{Tr}(\textbf{A}^2)/3$, the time
evolution of $\lambda$ is simply given by $d\lambda/dt = \lambda
^2$ and the finite time divergence is given by the solution
$\lambda(t) = \lambda(t_0)/(1-t\lambda(t_0)) $ in a finite time
$1/\lambda(t_0)$. In our case, we see from Eq.
(\ref{eq:InitVieille}) that the anisotropic part of the pressure
Hessian acts directly against the development of the singularity
induced by the self-streching term. Indeed, the coefficient in
front of $\lambda^2$ is bounded between $[-1;1]$. Thus this model
for pressure Hessian can regularize the finite time divergence
when $\lambda\gg \tau^{-1}$ since then the prefactor of
$\lambda^2$ is close to -1 and the solution of
$d\lambda/dt=-\lambda^2$ tends to zero at large times. Further
discussions of the regularization along the Vieillefosse tail
require specification of the time scale $\tau$, in particular its
dependence on the Reynolds number.

\section{Reynolds number effects, intermittency, and relative anomalous scaling}

In this section, three choices to model Reynolds number changes
are considered: (I) constant Kolmogorov time scale (this was the
case studied in \cite{CheMen06} and for clarity the relevant
results will be repeated here), (II) local time scale, and (III)
variable forcing strength.

Intermittency is studied by examining the scaling of moments of
velocity gradients. As in \cite{CheMen06} we consider both
longitudinal ($A_{ii}$, no index summation) and transverse
($A_{ij}$, $i \neq j$) gradients. Nelkin \cite{Nel90} shows,
assuming the relevance of the multifractal formalism in the
inertial range, that
the relative scaling of higher order moments of velocity
derivatives should behave as a power law
\begin{equation}\label{eq:Nelkin}
\left\langle \left| A_{ij} \right|^p\right\rangle \sim
\left\langle \left| A_{ij} \right|^2\right\rangle ^{ \mathcal
F^{{\tiny \mbox{L,T}}}(p)/2}\mbox{, with } \mathcal F^{{\tiny
\mbox{L,T}}}(p) = \min_h\left[ -\frac{p(h-1) + 1-\mathcal
D^{{\tiny \mbox{L,T}}}(h)}{h+1}\right]\mbox{ .}
\end{equation}
These equations are written for either longitudinal ($i=j$,
superscript $L$) or transverse ($i \neq j$, superscript $T$)
gradients. The functions $\mathcal D^L(h)$ and  $\mathcal D^T(h)$
are the longitudinal and transverse singularity spectrum,
respectively. Imposing $\zeta_3=1$ in the inertial range (which is
exact for longitudinal velocity increments, and a good
approximation for transverse ones) leads to $\langle
\left(A_{ij}\right)^2\rangle \sim \mathcal R_e$, i.e. finiteness
of dissipation is recovered \cite{Fri95}.

To proceed and facilitate interpretation of results, as in Ref.
\cite{CheMen06} we choose a simple quadratic form for the
singularity spectrum $\mathcal D^{L,T}(h) =
1-(h-c_1^{L,T})^2/(2c_2^{L,T})\mbox{, with
}c_1^{L,T}=\frac{1}{3}+\frac{3}{2}c_2^{L,T}$ where the parameter
$c_2^{L,T}$ is called the intermittency exponent (see Ref.
\cite{CheCas06} for further details).

{\it{Case I  -  Constant Kolmogorov time scale}:} First we
consider the simplest case in which $\tau$ is a constant value.
Because $\tau$ should scale with the Lagrangian decorrelation
time-scale of the velocity gradients,  it is chosen to be of the
order of the Kolmogorov time scale $\tau_K$ \cite{CheMen06}. This
choice gives an explicit Reynolds number dependence to the model
since it that case, $\tau_K  \sim \mathcal R_e^{-1/2}$.
As argued earlier already, from Eq. (\ref{eq:InitVieille}), we see that when
$\lambda \gg \tau^{-1}$, the model pressure Hessian causes the
coefficient appearing in front of $\lambda^2$ to switch from 1 to
-1. Thus, it acts to counteract the finite time divergence and
causes $\lambda$ to decrease in time as $1/t$. The viscous part is
also very important and can be seen as a very efficient damping
term with a coefficient which grows exponentially with increasing
$\lambda$'s. We have checked numerically that for any initial
conditions, the system Eq. (\ref{eq:InitVieille}) is such that
$\lambda(t)\rightarrow 0$ when $t\rightarrow +\infty$. The
divergence is thus regularized. We have also checked numerically
that for any other initial conditions for $\textbf{A}$, i.e. those
which cannot be written as in Eq. (\ref{eq:InitVieille}), all
components of $\textbf{A}$ evolving under Eq.
(\ref{eq:Determourmodel}) tend to zero in the absence of forcing.
Without loss of generality, henceforth all variables will be
scaled with the time-scale $T$, i.e.  $t/T \to t$ and $A_{ij} T\to
A_{ij}$. Let us denote by $\Gamma=\tau/T$ the only free parameter.
The forcing term $d\textbf{W}=\textbf{G}\sqrt{2dt}$ is Gaussian
and its covariance matrix is assumed to be Reynolds number
independent  (see Ref. \cite{CheMen06} for details). The model
(Eq. (\ref{eq:Determourmodel})) is solved numerically according to
Ref. \cite{CheMen06} and stationary statistics are obtained. The
results are examined from the point of view of intermittency and
anomalous relative scaling properties.

 \begin{figure}[t]
\center{\epsfig{file=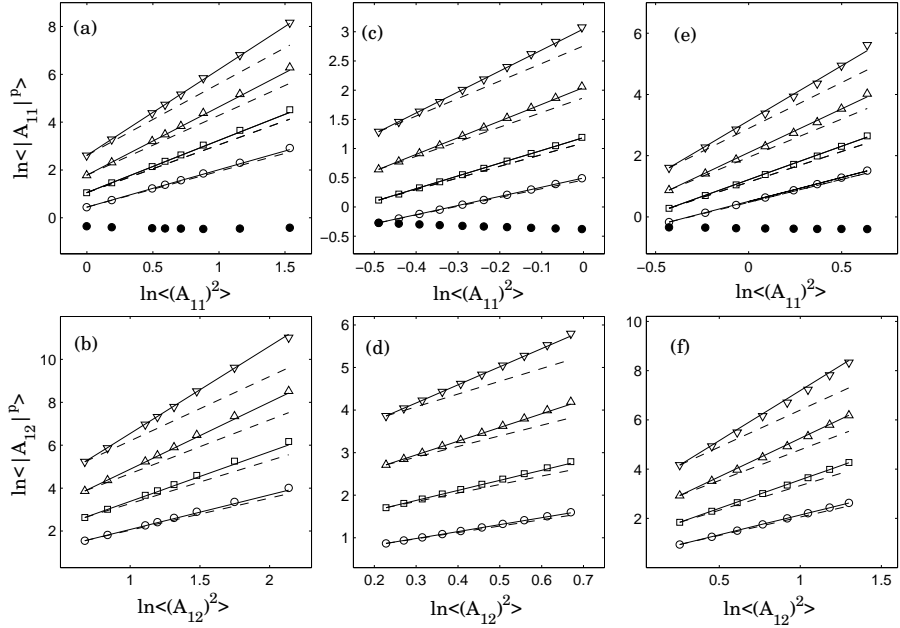,width=12cm}}
\caption{Relative scaling properties of higher order moments, i.e.
$\ln \langle  |A_{ij}|^p\rangle$, as a function of the second
order moment $\ln\langle  |A_{ij}|^2\rangle$. Dashed-line
represents K41 predictions (i.e. $p/2$). Solid lines: multifractal
predictions (i.e. $\mathcal F(p)/2$, see text), and $\circ$
($p=3$), $\square$ ($p=4$), $\bigtriangleup$ (p=5) and
$\bigtriangledown$ (p=6). Skewness of longitudinal components is
represented with $\bullet$. Case I (a-b): $\Gamma\equiv \tau_K /T$
=0.2, 0.15, 0.1, 0.09, 0.08, 0.07, 0.06 and 0.05 (same value as in
Ref. \cite{CheMen06}). Case II (c-d): Local time scale
$\Gamma/{\sqrt{2\mbox{Tr}(\textbf{S}^2)}}$ with $\Gamma=1.4; 1.35;
1.3; 1.25; 1.2; 1.15; 1.1; 1.05; 1$ and 0.95. Case III (e-f):
Local time scale $1/{\sqrt{2\mbox{Tr}(\textbf{S}^2)}}$ with
increasing variance of the forcing term $d\textbf{W} = \sigma
\textbf{G}\sqrt{2dt}$, with $4\sigma^2 = 3; 3.5; 4; 4.5; 5; 5.5;
6; 6.5$. All parameters listed correspond to points going from
left to right.} \label{fig:FigAnomT}
\end{figure}

To facilitate comparison with the cases considered in II and III,
in Fig. \ref{fig:FigAnomT}(a-b) we present one of the results of
Ref. \cite{CheMen06}. These results are obtained from numerical
integration of the model system over long periods of time and the
evaluation of moments of various orders. Clearly, intermittency is
predicted because the K41 line (i.e. of slope $p/2$, dashed
lines)) does not fit the computed results from the model. We
display (solid lines) various predictions obtained with the help
of Eq. (\ref{eq:Nelkin}). The results can be described well with
the parameters $c_2^L=0.025$ and $c_2^T=0.040$ for the
longitudinal and transverse cases, respectively. Transverse
gradients appear to be more intermittent than longitudinal ones.
Also, as shown in  Ref. \cite{CheMen06} longitudinal gradients PDF
is skewed. As was stressed in Ref. \cite{CheMen06}, the
intermittency parameters $c_2^L$  and $c_2^T$ are very close to
those obtained from experimental data, see Ref. \cite{CheCas06,DhrTsu97}.
As remarked in Ref. \cite{CheMen06}, however, for values of
$\tau/T$ smaller than about 0.05 (corresponding to a Taylor-based
Reynolds number of order 300 \cite{YeuPop06}) the predicted
statistics become unrealistic. Still, the fact that a model with
only 8 degrees of freedom derived directly from the Navier-Stokes
equations predicts realistic relative intermittency exponents
(albeit in a limited range of Reynolds numbers) is quite
remarkable. The results raise the question of how robust these
findings are with respect to other possible choices of the
time-scale and forcing strengths. This is considered in the next
two subsections.

{\it{Case II - Reynolds number dependent local time scale}: } It
has been hypothesized that the dissipative scale in turbulence is
not  constant but fluctuates due to the intermittency phenomenon
\cite{PalVul87}. Consistent with this notion of a local
fluctuating cutoff scale, here we choose $\tau(t) =
\Gamma(\mathcal R_e)/\sqrt{2\mbox{Tr}(\textbf{S}^2)}$ where
$\Gamma$ is a dimensionless parameter, and an unknown function of
the Reynolds number. Hence in this case, as opposed to Case I where $\tau$ was
held constant,  the time-scale fluctuates, being a function the variable velocity gradients.
When $\Gamma$ decreases for a fixed
\textbf{A}, the predicted pressure Hessian is closer to isotropy.
Then the system (\ref{eq:Determourmodel}) is  dominated more by
the quadratic (singularity-inducing) self-stretching term, which
is what one may expect at higher Reynolds numbers. In this case,
for a fluctuating dissipative time scale, the ODE appearing in
equation Eq. (\ref{eq:Determourmodel}) can be solved exactly and
one obtains
\begin{equation}\label{eq:LocGammTime}
\lambda(t) =
\frac{\lambda_0b^2}{3a\lambda_0-e^{\frac{b}{3}t}(-b^2+3a\lambda_0)}\mbox{
, where
}a=4e^{-\frac{2}{\sqrt{12}}\Gamma}-e^{\frac{4}{\sqrt{12}}\Gamma}
\mbox{ and }
b=2e^{-\frac{2}{\sqrt{12}}\Gamma}+e^{\frac{4}{\sqrt{12}}\Gamma}>0\mbox{
.}
\end{equation}
The long time behavior of the solution depends on the sign of the
constant $a$. For $\Gamma>\frac{\sqrt{12}}{6}\ln 4 \approx 0.80$,
$\lambda (t) \rightarrow 0$ at large times. For smaller
$\Gamma$'s, the solution diverges in a finite time (when
$\lambda_0>b^2/(3a)$) and the model is unable to regularize the
divergence predicted by the self streching term. Therefore,
similarly to the constant Kolmogorov time-scale option considered
in I, this approach appears not to allow reaching arbitrarily high
Reynolds numbers. The system using different values of $\Gamma$ is
integrated numerically as in I, with the same Gaussian forcing
$d\textbf{W} = \textbf{G}\sqrt{2dt}$. Resulting moments of
velocity gradients are displayed in Fig. \ref{fig:FigAnomT}(c-d).
As in I, anomalous relative scaling and intermittency is obtained.
The solid lines again are obtained by using a longitudinal
intermittency coefficients $c_2^L=0.025$, the same as the one
obtained with a constant time scale. The transverse coefficient is
$c_2^T=0.045$, also almost the same as that obtained in I.

{\it{Case III: Variable forcing strength}:} Another option to
model Reynolds number effects is to vary the strength of the
stochastic forcing through the term $d\textbf{W}=\sigma \textbf{G}
\sqrt{2dt}$. The variance $\sigma ^2$ is assumed to increase with
increasing Reynolds number based on the notion that the relative
strength of forcing (compared to viscous term) from neighbouring
and large eddies increases with Reynolds number. The explicit
dependence on Reynolds number is unknown but since we use relative
scaling, the precise relationship with Reynolds number is not
needed for the analysis. As a time-scale, in this case we use the
local time scale, namely $\tau(t)=
1/\sqrt{2\mbox{Tr}(\textbf{S}^2)}$,  i.e. the former time scale of
Eq. (\ref{eq:LocGammTime}) with $\Gamma=1$ (which is large enough
to insure regularization along the Vieillefosse tail). Numerically it is
observed that increased $\sigma$ leads to increased variance
of the velocity gradient components. This is consistent with the view that
changing the forcing may be viewed as modifying the effective
Reynolds number.

Numerical results for the moments of gradients and quantification
of relative scaling are presented in Fig. \ref{fig:FigAnomT}(e-f)
for various strengths $\sigma$ of the forcing. Once again, using
the representation of Eq. \ref{eq:Nelkin}, the results agree very
well with intermittent exponents consistent with $c_2^L=0.025$ and
$c_2^T=0.040$. Thus the quantitative predictions of anomalous
relative scaling in the model appear to be quite robust with
regard to details of the regularization time-scale and forcing.

\begin{figure}[t]
\center{\epsfig{file=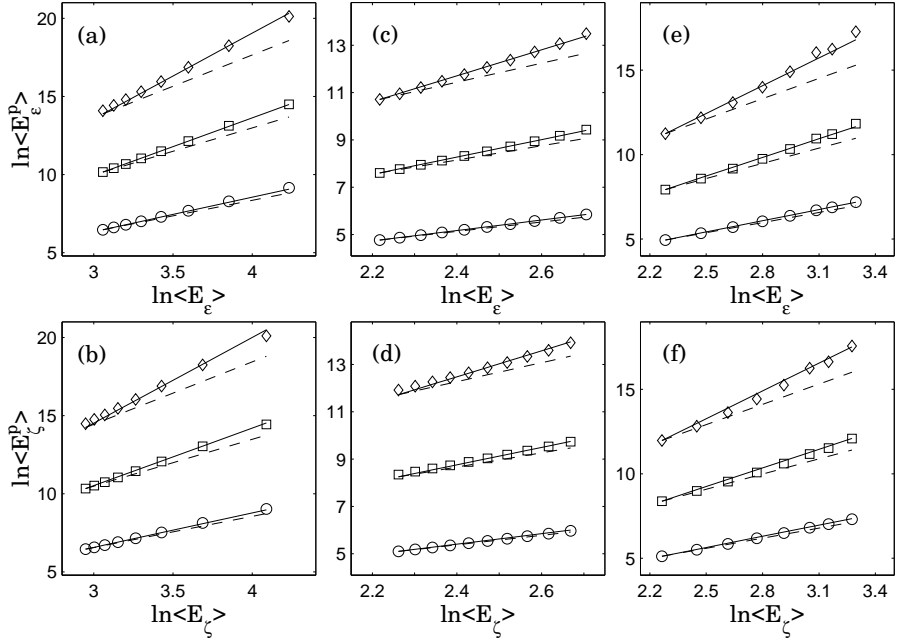,width=12cm}}
\caption{Relative scaling properties of higher order moments, i.e.
$\ln \langle E^p\rangle$, as a function of the average $\ln
\langle E\rangle$, where $E$ stands for ``dissipation''
$E_\epsilon$ or ``enstrophy'' $E_\zeta$. The cases (a)-(f)
correspond to the same model parameters as in Fig.
\ref{fig:FigAnomT}. Various orders of moments $\langle E^p\rangle$
are studied: $p=2$ ($\circ$), $p=3$ ($\square$) and $p=4$
($\diamond$). Dashed-line represents K41 predictions (i.e. slope
$p$), solid lines multifractal predictions (i.e. slope $p+\mathcal
F(p)$, see Ref. \cite{NotesMultiMeas}).
 } \label{fig:FigAnomMes}
\end{figure}

\section{Relative scaling properties of dissipation and enstrophy}

The model (Eq. (\ref{eq:Determourmodel})) can also be used to
predict the dynamics and statistics of several norms of the
velocity gradient tensor, namely $E_\epsilon =
2\mbox{Tr}(\textbf{S}^2)$  (the dissipation divided by viscosity),
$E_\zeta=2\mbox{Tr}(\vct{\Omega}\vct{\Omega}^T)$ (the enstrophy,
where $\vct{\Omega}$ is the antisymetric part of \textbf{A}) and
$E_\varphi=\mbox{Tr}(\textbf{A}\textbf{A}^T)$ (the
``pseudo-dissipation'' divided by viscosity). It is well known
that in turbulent flows, these ``dissipation fields'' are highly
intermittent \cite{Fri95,Meneveauetal90,MenSre91}. In homogeneous
and isotropic turbulence, $\langle E_\epsilon\rangle=\langle
E_\zeta\rangle=\langle E_\varphi\rangle$, while higher order
moments of these quantities are not similarly linked. We will
perform then, in a similar fashion as Fig. \ref{fig:FigAnomT}, a
relative scaling study of these quantities.

We present in Fig. \ref{fig:FigAnomMes}, similar to Fig.
\ref{fig:FigAnomT}, the numerical results for relative scaling  of
$E_\epsilon$ and $E_\zeta$ obtained from a numerical integration
of the model (Eq. (\ref{eq:Determourmodel})), for the three cases
I, II and III. Clearly, numerical results don't follow K41
predictions (dashed line). It is found that for the all three
cases the relative scaling properties are the same, i.e. again we
observe robustness with respect to how Reynolds number effects are
modeled. Finally, we observe that dissipation and enstrophy scale
the same. The solid line shows the multifractal predictions
\cite{NotesMultiMeas,SreMen88,MenNel89,Bor93},  using  a unique intermittency parameter
$\mu=0.25$. Similar results are obtained when studying the
relative scaling properties of the pseudo-dissipation
$E_{\varphi}$, i.e. we obtain $\mu \approx \mu^{\varphi} \approx
0.25$ (data not shown).  Thus, in the model the dissipation,
enstrophy and ``pseudo-dissipation'' display the same or very
similar intermittency exponents. The value of $\mu \approx 0.25$
is in excellent agreement with previous numerical and experimental
investigations \cite{Fri95,MenSre91}.

\section{Summary and Conclusions}

Several options to represent Reynolds number in a Lagrangian
velocity gradient model using the newly proposed RFD closure
\cite{CheMen06} have been considered. Numerical integrations of
the model show that three different ways to represent Reynolds
number variations have led to the same (universal) intermittency
relative scaling exponents. Consistent with data, it has been
found that longitudinal intermittency exponent is of order $c_2^L
= 0.025$ and the transverse one is $c_2^T = 0.040$. Relative
scaling analysis for dissipation, enstrophy and pseudo-dissipation
shows that all three quantities share the same intermittency
coefficient ($\mu \approx 0.25$) in the model predictions.  The
universality of the results highlights the importance of the
self-streching term in the dynamics of \textbf{A}. Future work
will be devoted to test the robustness of the predicted
intermittency phenomenon with respect to possible modifications of
this term.

At this stage it may be of interest to recall some prior
discussions dealing with the various scaling exponents in
turbulence. On the one hand, and in agreement with our model
results, numerically and experimentally the longitudinal and
transverse velocity gradient intermittencies were found to be
different (see for instance Ref. \cite{JiaGon06} for a recent
review on the subject). Conversely (and unlike our model's
results), they have been predicted to be the same from a field
theoretic approach (see Ref. \cite{BifPro05} and references
therein). On the other hand, unlike our model results, enstrophy
was found to be more intermittent than dissipation in numerical
flows \cite{CheSre97} (and in experiments, albeit using only a
single component of vorticity \cite{Meneveauetal90}), whereas (in
agreement with our model) they are predicted to scale the same at
infinite Reynolds numbers from simple arguments based on the
finiteness of the inertial range pressure spectrum \cite{Nel99},
or from more systematic irreducible group representations
\cite{BifPro05}. It is therefore interesting to note that the
present model provides, for a limited range of Reynolds numbers,
an example of dynamics of the velocity gradient tensor in which
transverse velocity gradients are more intermittent than
longitudinal ones, whereas dissipation and enstrophy scale the
same. Clearly more research is needed to elucidate the
relationships between various exponents characterizing
intermittency in turbulence, and possibly to clarify
generalizations of the refined similarity hypothesis (e.g. such as
in \cite{CheSre97}). The present approach of using Lagrangian
dynamical evolution equations \cite{CheMen06} should help shed new
light on this long-standing, important problem.

Acknowledgements: We thank M. Nelkin for motivating us to
examine scaling properties of dissipation and enstrophy, and Y.Li, Z.
Xiao, L. Biferale, S. Chen, G. Eyink and F. Toschi for useful
suggestions. L.C is supported by postdoctoral Fellowship from
the Keck Fundation and C.M. by the
National Science Foundation.


\begin{thebibliography}{00}
\bibitem{NotesMultiMeas}
Multifractal predictions for relative scaling properties of the
dissipation signals $E$ (either $E_\epsilon$, $E_\zeta$  or
$E_\varphi$) may be derived in analogy with Eq. (\ref{eq:Nelkin})
\cite{SreMen88,MenNel89,Bor93}: $\langle E^p\rangle \sim \langle
E\rangle ^{ p+\mathcal F(p)}\mbox{, with } \mathcal F(p) =
\min_\alpha\left[ -3 \left(p(\alpha-1) + 3-
f(\alpha)\right)/(\alpha+3)\right]$. The singularity spectrum
$f(\alpha)$ is such that $\mathcal F(p)$ vanishes for both $p=0$
and $p=1$. Assuming furthermore a simple parabolic form for
$f(\alpha)$, we are left with a unique parameter $\mu$ usually
called the intermittency parameter in the literature \cite{Fri95}:
$f(\alpha) = 3- [\alpha- (1+\mu)]^2/(2\mu)$. The set of exponents
$\mathcal F(p)$, $f(\alpha)$, and $\mu$ may in principle differ
for each of the fields $E_\epsilon$, $E_\zeta$  and $E_\varphi$.
\end{thebibliography}
\end{document}